\documentclass[showpacs,prl,twocolumn,aps]{revtex4}

\usepackage{graphicx}

\begin{document}
%%%%%%
\title{Top anti-top pairs at the LHC heavy ion collision: a new interesting 
probe of quark gluon plasma}
%%%%%%
\author{Lusaka Bhattacharya$^1$, Kirtiman Ghosh$^2$ and Katri Huitu $^3$}
\affiliation{1. Saha Institute of Nuclear Physics,  Bidhannagar, India.\\
lusaka.bhattacharya@saha.ac.in\\
2. \& 3. Department of Physics and Helsinki Institute of Physics\\
FIN-00014, University of Helsinki, Finland.\\
kirti.gh@gmail.com, Katri.Huitu@helsinki.fi}
%%%%%
\date{\today}
%\maketitle

\begin{abstract}
We investigate top anti-top quark pair production in lead-lead collisions  at the Large Hadron Collider with nucleon-nucleon center of mass energy of 5.5 TeV. Due to the very high temperature and energy density created in heavy ion collision, a new state of QCD matter known as Quark-Gluon Plasma (QGP) is expected to be produced. Top decay products loose energy inside the QGP medium. Therefore, we also study the medium modifications of different kinematic distributions. We observe significant modification in the dijets and trijets invariant mass distributions.We also found that the peak position and shape of the distributions could be used to  characterize the nature of jet energy loss in the QGP.

\end{abstract}

%\
%\pacs{}

\maketitle

The primary goal of heavy ion collisions (HIC) at Relativistic Heavy Ion Collider (RHIC) at BNL, and the Large Hadron Collider 
(LHC) at CERN is to produce and study the properties of a hot/dense state of QCD matter known as {\em Quark-Gluon Plasma} 
(QGP)~\cite{dk1}. QGP is a deconfined state of matter where quarks and gluons are the effective degrees of freedom rather 
than nucleons or hadrons~\cite{dk2}. Fast partons propagating in a hot/dense nuclear medium are expected to loose a large 
fraction of their energy \cite{jq}. The observation of the suppression of energetic partons in the QGP, that is {\em jet quenching} \cite{jq_RHIC}, 
and centrality-dependent {\em dijet asymmetry} \cite{da_LHC} are the most important results from the HIC at RHIC and LHC experiment, respectively.

$W/Z$-bosons are massive weakly interacting  Standard Model (SM) particles. 
%They have been extensively studied at CERN, SLAC and FNAL 
%in $p\bar p$ and $e^+ e^-$ collisions 
%and considered as the Standard Model (SM) benchmarks. 
The larger LHC HIC energies 
open the possibility to probe the nucleus-nucleus collisions via the $W/Z$-bosons. The vector bosons are produced early ($1/M_{W(Z)} \sim 10^{-3}$ fm/c) and  their decay time is small ($\tau_Z \sim 0.08$ fm/c and $\tau_W \sim 0.09$ fm/c \cite{pdg}). 
Whereas, in the most accepted picture of the QGP formation and evolution, 
at the LHC, QGP is expected to form after 
$1/\Lambda_{\rm QCD}\sim 1$ fm/c of the initial hard scattering, thermalize quickly and it might last $\sim 10$ fm/c. 
Therefore, weak bosons are produced and decay before the formation of QGP and the decay products of pass through the QGP. $W/Z$-boson dominantly decays to a pair of 
quarks.
% with branching fraction (BF) $67.6\% (69.9\%)$. 
Quarks loose energy in the QGP and thus the hadronic decays of $W/Z$-boson could be an interesting probe to characterize QGP. 
However, in presence of huge QCD dijet background, it is extremely challenging to study the hadronic decays of  $W/Z$-bosons. The leptonic BF of $W/Z$-boson is small however, due to small background, the signature could 
be easily detected at the LHC HIC.
% $BF(W\to l \nu)=10.8\%$ 
%and $BF(Z\to l^+ l^-)=3.4\%$. 
%However, the leptonic decay channels are less affected by the background and thus could 
%be easily detected at the LHC HIC. 
Unlike jets, leptons interact electromagnetically with the QGP and loose
experimentally insignificant amount of energy within the QGP \cite{ConesadelValle:2009vp}. Therefore, the leptonic decays of weak bosons behave 
as a medium blind reference.
% and cannot be used to characterize the properties of QGP. 
%However, weak bosons provide 
%a direct measurement of the primary nucleon-nucleon binary collisions and this can be used to normalize the production 
%mechanisms of various other processes. 
ATLAS collaboration has already measured the $W/Z$-boson yield in the 
leptonic decay channels for $\sqrt {s_{NN}}=2.76$ TeV \cite{ATLAS_WZ}.
% and integrated luminosity of $0.15$ nb$^{-1}$ of Pb+Pb collision.  

A new regime of heavy ion physics will be reached at the LHC with $\sqrt {s_{NN}} = 5.5$ TeV where hard and semi-hard 
particle production can dominate over the underlying soft events. The higher LHC energies open the possibility to 
study top quarks for the first time at the HIC. The top quark was discovered at the Tevatron 
experiment \cite{top}. After the discovery, properties of top quark have been extensively studied at the Tevatron and LHC. Presently, different properties of top quark is known with good precision. 
As an example, top quark mass and full decay width are determined to be $m_{t}=173.5\pm 0.6$ GeV and $\Gamma_t=2.0^{+0.7}_{-0.6}$ GeV \cite{pdg}. 
Different decay channels and branching ratios of top quark have also been observed and measured.
% at the Tevatron and LHC 
%experiment. 
As a result, top quark can now be considered as standard benchmark for other experimental observations. In 
view of this fact, it is important to investigate top quarks at the LHC HIC and study the medium (QGP) influence on 
different kinematic distributions which are precisely known from the previous $pp$ and $p\bar p$ collider experiments. 
In this letter, we have for the first time studied $t\bar t$ production at the LHC HIC and proposed few kinematic 
distributions for the study of the QGP created in the LHC HIC.

The main source of top quarks at the LHC HIC is the top anti-top ($t \bar t$) pair production.
% via the strong interaction. 
At leading order in perturbation theory there are two processes that contribute to $t\bar t$ production: 
quark-antiquark annihilation, $q\bar q \to t\bar t$ and gluon-gluon fusion, $gg \to t\bar t$. With $5.5$ TeV center-of-mass 
energy per nucleon, the NLO+NNLL $t\bar t$ production cross-section per nucleon-nucleon collision in Pb-Pb reaction is 
given by $\sigma_{NN}(t\bar t)=80.6$ pb \cite{top_cross}. Therefore, the total $t\bar t$ production cross-section in minimum bias Pb-Pb 
scattering is estimated to be 3.5 $\mu b$ in the frame work of the Glauber model \cite{GM}. The instantaneous luminosity of the LHC 
Pb-Pb collision at $\sqrt s_{NN}=5.5$ TeV is expected to be $10^{27}{\rm cm^2 s^{-1}}$. Therefore, $1$ nb$^{-1}$ integrated 
luminosity data will be accumulated within one month ($10^6$ second) of Pb-Pb collision at $\sqrt s_{NN}=5.5$ TeV. These 
estimations indicate that about 3500 $t\bar t$ events are expected to be produced in one month of LHC HIC running at $\sqrt s_{NN}=5.5$ TeV.

Due to the large top mass, $t \bar t$ pairs are produced early in the HIC. The decay width of top quark is large and 
thus the decay time is small. As a result, top quarks are produced and decay before formation of QGP. 
Top quark decays to a bottom quark ($b$) and $W$-boson with almost 100\% branching fraction: $t\to b W^+$. $W$-boson subsequently 
decays hadronically ($W^\pm \to q \bar q^{\prime}$) with  67.7\% BF or leptonically ($W^\pm \to l \nu_l$) with $32.3$\% BF \cite{pdg}. Therefore, for $t \bar t$ production, there are only three possible final state topologies:
%
%\begin{itemize}
{\em (i) Hadronically decaying $t \bar t$ pairs:} Both the top quarks decay hadronically ($t\to b q \bar q^{\prime}$) 
and give rise to 2-$b$ jets and four light quarks jets in the final state with 46\% effective branching ratio.
{\em (ii) Semi-leptonically decaying $t \bar t$ pairs:} One top quark decays hadronically and the other decays leptonically 
($t\to b l \nu_l$). The final state is characterized by $2$-$b$ jets+$2$-light quark jets+one charged lepton + one neutrino. 
The effective branching fraction of semi-leptonic decay mode of $t \bar t$ pairs is $43.7$\%.
{\em (iii) Leptonically decaying $t \bar t$ pairs:} In this case, both the top quarks decay leptonically giving rise to 
$2$-$b$ jets + $2$-lepton + $2$-neutrino final state with 10.4\% effective branching fraction. 
%\end{itemize} 

The jets ($b$-jets as well as light jets) and leptons can be observed at the LHC. The neutrinos remain invisible at the 
detectors and give rise to a imbalance in the transverse momentum known as missing transverse momentum.  However, in the 
HIC environment, faithful measurement of missing transverse momentum will be challenging in the presence of a continuum 
energy deposit from the produced QGP. 
%At the LHC HIC with $\sqrt s_{NN}=5.5$ TeV and $1$ nb$^{-1}$ integrated luminosity, 
%we will have about $1600$ hadronically decaying $t \bar t$ events,  $1500$ semi-leptonically decaying $t \bar t$ events and 
%$365$ leptonically decaying $t \bar t$ events.

Hadronically decaying $t \bar t$ events suffer from huge QCD background. Moreover, due to a large combinatorial background 
it is difficult to reconstruct top quark and $W$-boson from hadronically decaying $t\bar t$ pairs. Leptonically 
decaying $t \bar t$ channel is a clean signal channel due to less background. However, the rate of this channel is 
suppressed by the top quark leptonic branching fraction. Moreover, in the absence of proper knowledge about missing transverse 
momentum, the reconstruction of top quark and $W$-boson mass will be difficult from $t \bar t$ leptonic decay channel. 
As a result, in this letter, we have investigated the semi-leptonic decay channel of $t \bar t$ pairs due to the following 
advantages: (i) The rate of semi-leptonic $t \bar t$ final state is relatively large. If we consider only electron and muon 
decay modes, Pb-Pb collision with $\sqrt s_{NN}=5.5$ TeV gives rise to about 1000 semi-leptonically 
decaying $t \bar t$ events for $1$ nb$^{-1}$ integrated luminosity. (ii) Due to the presence of a lepton, the semi-leptonic 
final state of $t\bar t$ pairs suffers less from the QCD background. (iii) At the parton level, the semi-leptonic 
$t\bar t$ final state contains only $2$-light quark jets arising from the $W$-decay. Therefore, $W$ invariant mass can be 
constructed with out any ambiguity. However, there is a two fold ambiguity in the reconstruction of top quark mass. 

After the decay of $t \bar t$ pairs, the decay products pass through the hot and dense QGP medium and thus, loose energy. The energy loss of energetic partons, so called "jet quenching", leads to a number of
phenomena which are already seen at RHIC and LHC. In this work, we have investigated the quenching of $t \bar t$ decay products and proposed some new phenomena which could be seen at the LHC HIC. We have used PYTHIA \cite{pythia} to simulate the production and decay of $t \bar t$ pairs. Subsequently, the PYTHIA generated $t \bar t$ events are passed in to a fast Monte-Carlo simulator PYQUEN \cite{pyquen} for simulating the energy loss (quenching) of $t \bar t$ decay products. Finally, quarks and gluons are hadronized according to the Lund string mode \. PYQUEN simulates the radiative and collisional (Coll.) energy loss \cite{coll_loss} of hard partons in longitudinally expanding QGP taking into account the realistic nuclear geometry. The radiative energy loss is calculated in the framework of BDMS model \cite{bdms} with the simple generalization to a massive quark case using the “dead-cone” approximation \cite{dead-cone}. 
%The radiative energy losses of a high energy parton have been shown to dominate over the collisional losses by up to an order of magnitude \cite{dominate}. However, a direct experimental verification of this phenomenon remains an open problem. Moreover, 
Measuring jet energy as a sum of the energies of final hadrons moving inside an angular cone with a given finite size allow some of the radiated gluons to belong to the jet and thus some part of the radiated energy to be reconstructed. Therefore, the knowledge of angular structure of medium-induced radiation is very important for any phenomenological prediction using jets. In this analysis, we have used the simple parametrizations of the gluon distribution over the emission angle $\theta$ available in PYQUEN: (i) Small-angular radiation (SAR): $dN^g/d\theta \propto {\rm sin}\theta {\rm exp}[-(\theta-\theta_0)^2/2\theta_0^2]$, where $\theta_0\sim 5^0$ is the typical angle of the coherent gluon radiation as estimated in Ref.~\cite{SA}; (ii) Wide-angular radiation (WAR): $dN^g/d\theta \propto 1/\theta$. The strength of the energy loss in PYQUEN is determined mainly by the initial maximal temperature $T_0^{max}$ of hot matter in Pb-Pb collisions. The energy loss also depends on the number $N_f$ of active flavors in the medium. In our analysis, we have used $T_0^{max}=1$ GeV and $N_f=2$.

%%%%%%%%%%%%%%%%%%%%%%%%%%%%%%%%%%%%%%%%%%%%%%%%%%%%%%%%%%%%%%
 \begin{figure}
 \includegraphics[angle=-90,width=80mm]{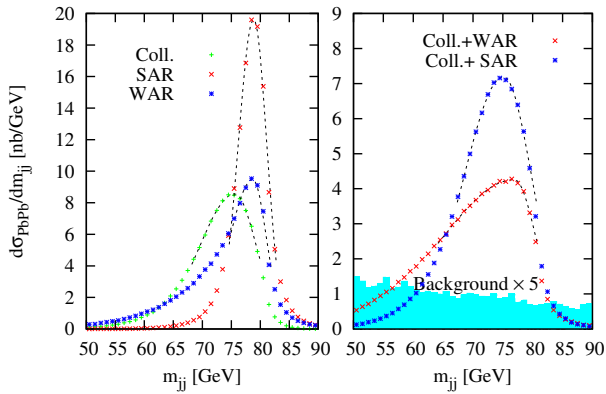}
 \caption{Invariant mass distribution of hardest and second hardest non $b$-jets for different energy loss scenarios. To determine the peak position of the distribution, we have fitted these distributions with asymmetric Gaussian functions. The parameters of fitting are presented in Table.~\ref{fit}}
\label{wmass}
 \end{figure}
%%%%%%%%%%%%%%%%%%%%%%%%%%%%%%%%%%%%%%%%%%%%%%%%%%%%%%%%%%%%%%

%%%%%%%%%%%%%%%%%%%%%%%%%%%%%%%%%%%%%%%%%%%%%%%%%%%%%%%%%%%%%%
 \begin{figure}
 \includegraphics[angle=-90,width=80mm]{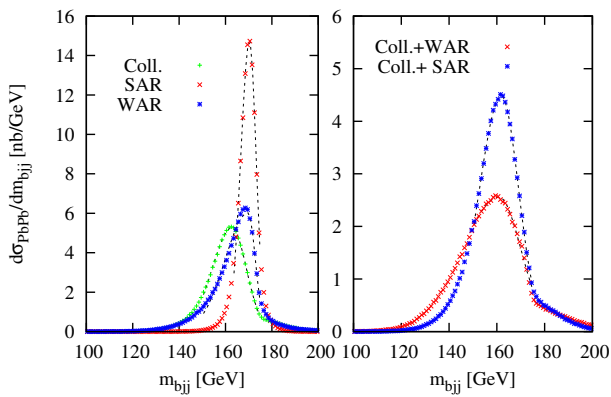}
 \caption{Same as Fig.~\ref{wmass} but for $m_{bjj}$ distribution.}
\label{tmass}
 \end{figure}
%%%%%%%%%%%%%%%%%%%%%%%%%%%%%%%%%%%%%%%%%%%%%%%%%%%%%%%%%%%%%%

In this letter, we have investigated $4$-jets out of which $2$-jets are $b$-tagged plus one lepton signature as signal of 
$t\bar t$ production in HIC. Therefore, before going into the details of our analysis, it is important to discuss the 
status of jet reconstruction and $b$-tagging in the context of HIC. 
%Experimental measurements so far have mostly been
%restricted to leading particle suppression relative to $pp$ collisions because of the limited center-of-mass energies
%available. However, in order to constrain the underlying QCD theory of jet quenching,
%new and more differential observables are needed. 
The main obstacle to studying jets in HIC is the presence of the huge background given by the underlying event (UE) 
%produced simultaneously with the hard nucleon-nucleon collision that
%initiates the high-transverse momentum jet of interest. 
This UE needs to be properly subtracted
from the momentum of a given jet in order to reconstruct its “true” momentum. It was shown in Ref.~\cite{Cacciari:2010te} that in presence of QGP, faithful reconstruction of jets are possible using anti-$k_T$ algorithm. Jets have already been successfully used as an observable in HIC by the ATLAS collaboration~\cite{da_LHC}. Moreover, in Ref.~\cite{vitev}, different jet shape variables have been studied for the jets passing through QGP. In Ref~\cite{White:2005au}, $b$-tagging in the environment of HIC has been studied by examining the reconstruction efficiency and rejection power (against light quark jets) using secondary vertex finding. Their study suggests that $\epsilon_b=50\%$ $b$-tagging  efficiency can be achieved at the HIC for a rejection power of 50. 
%In our analysis, we have used anti-$k_T$ algorithm with $R=\sqrt{\Delta \eta^2+\Delta \phi^2}=0.4$ for the jet reconstruction and $b$-tagging efficiency of 50\%.

%%%%%%%%%%%%%%%%%%%%%%%%%%%%%%%%%%%%%%%%%%%%%%%%%%%%%%%%%%%%%%%%%%
 \begin{table}%[H] add [H] placement to break table across pages
\vskip -15pt
 \caption{Parameters used for the Gaussian fitting of Fig~\ref{wmass}~and~\ref{tmass}.}
\label{fit}
 \begin{ruledtabular}
\begin{tabular}{||c||c|c|c|c||c|c|c|c||}
Energy loss & \multicolumn{4}{c||}{For $m_{jj}$ distribution} & \multicolumn{4}{c||}{For $m_{bjj}$ distribution} \\\cline{2-9}
Scenario  & $a_0$ & $m_0$ & $\sigma_1$ & $\sigma_2$ & $a_0$ & $m_0$ & $\sigma_1$ & $\sigma_2$ \\
    &  nb & GeV & GeV & GeV &  nb & GeV & GeV & GeV\\\hline \hline
Only Coll. & 8.6 & 75.6 & 6.1 & 3.8 & 5.3 & 162.8 & 8.6 & 6.2\\
Only SAR & 19.6 & 78.9 & 2.6 & 2.2 & 14.8 & 170.4 & 3.7 & 2.8 \\
Only WAR & 9.5 & 78.7 & 4.0 & 2.4 & 6.2 & 169.9 & 9.9 & 3.0 \\\hline 
Coll.+SAR & 7.16 & 75 & 6.4 & 4.9 & 4.5 & 161.9 & 8.9 & 7.5 \\
Coll.+WAR & 4.25  & 76.9 & 12.1 & 3.3 &  2.6 & 160.9 & 14.9 & 9.25 \\
\end{tabular}
 \end{ruledtabular}
\vskip -15pt
 \end{table}
%%%%%%%%%%%%%%%%%%%%%%%%%%%%%%%%%%%%%%%%%%%%%%%%%%%%%%%%%%%%%%%%%%%%%

In our analysis, we have introduced a set of basic selection criteria to identify electrons, muons, jets etc. The object selection is described in brief in the following: (i) Jets are constructed using anti-$k_T$ algorithm with $R=0.4$ and only jets with $p_T > 20$ GeV and $|\eta| < 2.5$ are considered for further analysis. To take into account the effects of finite detector resolution, we have smeared the jet energies with Gaussian functions in Ref.~\cite{smear}. (ii) We demand that lepton candidates (both electron and muon) have $p_T > 20$ GeV and are separated from jets by at least $\Delta R = 0.5$. After reconstructing different objects, we consider events with one lepton (electron or muon) and $\ge$ 4-jets for further analysis. We also demand that out of the 4-jets, two jets are $b$-tagged. The dominant background for semileptonic $t \bar t$ signal arises from $W/Z$+jets production followed by the leptonic decay of $W/Z$-boson. Here, $b$-jets results from the mistagging of light jets. Since the mistagging efficiency of light quark jets to be tagged as $b$-jets is small, $b$-tagging significantly reduce this background. Production of $W/Z~b\bar b$+jets also contributes to the background. However, we have estimated that these cross-sections are very small compared to the $t\bar t$ cross-section. Therefore, semileptonic decay products of $t\bar t$ could be easily detected over the background. 

To study the influence of the hot/dense QGP, we have constructed the following kinematic distributions.\\
{\em Di-jets invariant mass distribution:} We have ordered the non $b$ jets according to their $p_T$ hardness ($p_T^{j_1}>p_T^{j_2}>...$) and constructed the invariant mass of the hardest ($j_1$) and the second hardest ($j_2$) jet, $m_{jj}$. In the semileptonic decay of $t\bar t$ pairs, non $b$-jets arise from the decay of one $W$-boson. Therefore, in absence of QGP, $m_{jj}$ distribution should be peaked at the $W$-mass, $m_W=80.4$ GeV. However, in presence of QGP, $W$-decay product suffers energy loss.
% and thus the peak position of $m_{jj}$ distribution shifts from $m_W$. 
In Fig.~\ref{wmass}, we have presented the $m_{jj}$ distributions for different energy loss scenarios. To determine the peak position of the distributions, we have fitted the distributions with asymmetric Gaussian functions:\\
$$
f(m) = a_0 \left\{ \begin{array}{rl}
 {\rm exp}\left (-\frac{(m-m_0)^2}{\sigma_1}\right) &\mbox{ if $m<m_0$} \\
 {\rm exp}\left (-\frac{(m-m_0)^2}{\sigma_2}\right )  &\mbox{ otherwise}
       \end{array} \right.
$$
where, $a_0$, $m_0$ and $\sigma_{1,2}$ are the parameters of fitting. In Fig.~\ref{wmass} (left panel), we have presented $m_{jj}$ distributions for collisional and radiative (SAR as well as WAR) energy loss scenario separately. Corresponding fitting parameters are presented in Table~\ref{fit}. Fig.~\ref{wmass} (left panel) and Table~\ref{fit} show that collisional energy loss significantly (about 5 GeV) shifts the peak position of $m_{jj}$ distribution from $W$-mass. Whereas, radiative energy loss (both SAR and WAR) gives rise to small change in the peak position of $m_{jj}$ distribution. Due to the high boost of hard partons, the radiated gluons shift towards the parent parton direction and thus, resulting jets include large part of radiated gluon energies. As a result, radiative energy loss of hard partons has negligible impact on the peak position of $m_{jj}$ distribution. However, Fig.~\ref{wmass} (left panel) shows that WAR significantly changes the shape (which could be quantified by the fitting parameters $a_0,~\sigma_1~{\rm and}~\sigma_2$ in Table~\ref{fit}) of $m_{jj}$ distribution. In the passage of a fast parton through QGP, collisional and radiative energy loss occur simultaneously. Therefore, in Fig.~\ref{wmass} (right panel), we have presented $m_{jj}$ distribution in presence of both collisional and radiative energy loss. Corresponding fitting parameters are presented in the last two rows of Table~\ref{fit}. Table~\ref{fit} shows that the peak position of $m_{jj}$ distribution is determined by the collisional energy loss. Whereas, the shape parameters are governed by the nature of radiative loss. In Fig.~\ref{wmass} (right panel), we have also presented the background. The background contributions are substantially small compared to the $t \bar t$ contribution. \\
{\em Tri-jets ($bjj$) invariant mass distribution:} 
%We are considering events with two $b$-jets, two light jets and one lepton. 
Hadronic decay of one top quark gives rise to one $b$-jet and two light quark jets. Therefore, it is important to study invariant mass distribution of $b$ jet and di-jets system arising from the hadronic top quark decay. However, it is difficult to identify the $b$ jet and di-jets arising from the same top decay. There are several algorithm available in the literature for the reconstruction of top quark. However, most of the algorithms rely on the knowledge of $m_{bjj}$ peak position (top mass: $m_t=173.2$ GeV). In presence of QGP, energy loss of top decay products shift the peak position of $m_{bjj}$ distribution. Therefore, most of the top reconstruction algorithm are not applicable for Pb-Pb collision in their present form. In our analysis, we have used the following simplified algorithm for reconstructing  $m_{bjj}$ peak position. We first order $b$ jets according to their $p_T$ hardness ($p_T^{b_1}>p_T^{b_2}$) and constructed two invariant masses: $m_{b_1 jj}~{\rm and}~m_{b_2 jj}$. Out of these two invariant masses, we consider the invariant mass which is close to the $m_t=173.2$ GeV for plotting $m_{bjj}$ distribution. In Fig.~\ref{tmass}, we have presented $m_{bjj}$ distributions. In Fig.~\ref{tmass} (left panel), we have shown the effect of collisional and radiative energy loss on $m_{bjj}$ distribution separately. Whereas, right panel of Fig.~\ref{tmass} shows the resulting $m_{bjj}$ distribution if we consider both collisional and radiative energy loss simultaneously. We have also fitted these distributions with asymmetric Gaussian functions and the fitting parameters are presented in Table~\ref{fit}. Due to the energy loss, the position of $m_{bjj}$ peak shifts about 12 GeV from $m_t$. The shift could be easily observed at the LHC as a signature of QGP.  

To summarize, we have investigated semileptonic $t \bar t$ signature at the LHC HIC. Semileptonic $t \bar t$ signature could be easily observed at the LHC HIC with $\sqrt s_{NN}=5.5$ TeV. However, due to the presence of QGP $t \bar t$ decay products suffer energy loss and thus shape of different kinematic distributions are modified significantly. As for example, we have studied dijets and trijets invariant mass distributions and predicted significant change and shift in the shape and peak position of these distributions, respectively.

\end{document}